\documentclass[
twocolumn,
preprintnumbers,
nofootinbib,
prd,aps
]{revtex4}

\usepackage{epsf}

\begin{document}

\newcommand{\rem}[1]{{\bf #1}}
\renewcommand{\topfraction}{0.8}

\preprint{TU-829, UT-HET 016}

\title{
Cosmic-Ray Positron from Superparticle Dark Matter 
and the PAMELA Anomaly
}

\author{
$^{(a)}$Koji Ishiwata, $^{(b)}$Shigeki Matsumoto and $^{(a)}$Takeo Moroi
}

\affiliation{
$^{(a)}$Department of Physics, Tohoku University, 
Sendai 980-8578, Japan\\
$^{(b)}$Department of Physics, University of Toyama, 
Toyama 930-8555, Japan
}

\date{November, 2008}

\begin{abstract}

 Motivated by the anomalous positron flux recently reported by the
 PAMELA collaboration, we study the cosmic-ray positron produced by
 the pair annihilation and the decay of superparticle dark matter.
 We calculate the cosmic-ray positron flux and discuss implications
 of the PAMELA data.  We show that the positron excess observed by
 the PAMELA can be explained with some of the superparticle dark
 matter.

\end{abstract}

\maketitle

Dark matter, which accounts for about $23\ \%$ of the mass density of
the present universe \cite{Komatsu:2008hk}, has been a mystery in the
fields of particle physics, astrophysics, and cosmology.  Even though
there exist various well-motivated candidates for dark matter, like the
lightest superparticle (LSP), axion, and the lightest Kaluza-Klein
particle in the universal extra-dimension scenario, the nature of dark
matter is completely unknown and many possibilities have been
discussed to study dark-matter properties with direct and indirect
detections of dark-matter signals as well as with colliders.  Among
them, measurements of the anti-particle fluxes in the cosmic ray give
important information about dark matter; annihilation or decay of some
classes of dark matter particle produces energetic anti-particle,
which may be observed in the cosmic ray.

Very recently, the PAMELA collaboration has announced the first result
of the measurement of the positron flux for the energy range $1.5-100\
{\rm GeV}$ \cite{PAMELA}, which indicates an anomalous excess of the
positron flux over the expected background.  In particular, the PAMELA
results show a significant increase of the positron fraction at the
energy range $20\ {\rm GeV}\lesssim E\lesssim 100\ {\rm GeV}$, even
though the positron fraction is expected to decrease as energy
increases \cite{Moskalenko:1997gh,Baltz:1998xv}.  In addition, in the
past measurement of the positron flux by HEAT \cite{Barwick:1997ig},
similar excess was also pointed out.  To understand the anomalous
behavior of the positron flux, an unaccounted mechanism is
necessary from particle-physics and/or astrophysical point of view.

One important possibility of realizing enhanced cosmic-ray positron
flux is the annihilation or the decay of dark matter particle
\cite{Cirelli:2008pk,Barger:2008su}.  (Recently, another possibility
is also pointed out that the anomalous flux may be due to the positron
emission from Pulsars \cite{Hooper:2008kg}.)  In this study, we pursue
a possibility that the anomalous cosmic-ray positron is from the pair
annihilation or the decay of the dark matter particle.  There are
various candidates for dark matter.  Among those, in this study, we
consider one of the most well-motivated candidates for dark matter,
i.e., the LSP in models with low-energy supersymmetry (SUSY).
Conventionally, in the LSP dark matter scenario, it is assumed that
the $R$-parity is conserved and that the LSP is Bino-like lightest
neutralino or the gravitino.  Then, the cosmic-ray positron flux is
negligibly small.  This is because, in such a case, the positron is
produced by the pair-annihilation of the LSP, whose cross section is
significantly suppressed in the cases of the Bino-like and gravitino
dark matter.  However, the above assumptions are not necessary in
realizing the LSP dark matter scenario; in several well-motivated
scenarios, one of the above assumptions are relaxed and an enhanced
positron flux may be realized.

In this letter, we consider the implication of the positron-flux
observation of the PAMELA experiment to the SUSY model in which the
LSP becomes dark matter.  We calculate the production rate of the
cosmic-ray positrons in several scenarios.  Then, we show that, even
if we adopt the conventional estimation of the background $e^\pm$
fluxes, the positron excess observed by the PAMELA can be explained in
some of the cases.

Let us first summarize our procedure to calculate the positron flux
$\Phi_{e^+}$ (as well as the electron flux $\Phi_{e^-}$).  (For
detail, see \cite{Ishiwata:2008cu}.)  We solve the diffusion equation
to take account of the effects of the propagation of the positron.
The energy spectrum of the positron from dark matter
$f_{e^+}(E,\vec{r})$ evolves as \cite{Baltz:1998xv}
\begin{eqnarray}
 \frac{\partial  f_{e^+}}{\partial t}
 = K(E) \nabla^2 f_{e^+}
 + \frac{\partial}{\partial E}\left[ b(E) f_{e^+} \right]
 + Q.
 \label{diffeq}
\end{eqnarray}
The function $K$ is expressed as $K=K_0 E_{\rm GeV}^\delta$
\cite{Delahaye:2007fr}, where $E_{\rm GeV}$ is the energy in units of
GeV.  In our numerical calculation, we use the following three sets of
the model parameters, called MED, M1, and M2 models, which are defined
as $(\delta, K_0[{\rm kpc^2/Myr}], L[{\rm kpc}])=(0.70,0.0112,4)$
(MED), $(0.46,0.0765,15)$ (M1), and $(0.55,0.00595,1)$ (M2), with
$R=20\ {\rm kpc}$ for all models.  Here, $L$ and $R$ are the
half-height and the radius of the diffusion zone, respectively.  The
MED model is the best-fit to the boron-to-carbon ratio analysis, while
the maximal and minimal positron fractions for $E\gtrsim 10\ {\rm
 GeV}$ are expected to be estimated with M1 and M2 models,
respectively.  We found that the MED and M1 models give similar
positron fraction, so only the results with the MED and M2 models are
shown in the following.  In addition, we use $b=1.0\times
10^{-16}\times E_{\rm GeV}^2\ {\rm GeV/sec}$.  For the case where the
primary positron is produced by the pair annihilation of dark matter,
the source term is given by
\begin{eqnarray}
 Q_{\rm ann} = \frac{1}{2} B \langle \sigma v\rangle
 \frac{\rho_{\rm halo}^2(\vec{r})}{m_{\rm DM}^2}
 \left[ \frac{dN_{e^+}}{dE} \right]_{\rm ann}.
\end{eqnarray}
Here, $B$ is the so-called boost factor, $\langle\sigma v\rangle$ is
the averaged pair annihilation cross section, and $m_{\rm DM}$ is the
mass of dark matter particle.  In addition, $\rho_{\rm halo}$ is the
dark matter mass density for which we adopt the isothermal profile
\cite{IsoThermal} $\rho(\vec{r})=\rho_\odot (r_{\rm core}^2 +
r_\odot^2)/(r_{\rm core}^2+r_{\rm kpc}^2)$, where $\rho_\odot\simeq
0.43\ {\rm GeV/cm^3}$ is the local halo density, $r_{\rm core}\simeq
2.8\ {\rm kpc}$ is the core radius, $r_\odot\simeq 8.5\ {\rm kpc}$ is
the distance between the galactic center and the solar system, and
$r_{\rm kpc}$ is the distance from the galactic center in units of
kpc.  If the positron is from the decay of dark matter,
\begin{eqnarray}
 Q_{\rm dec} = \frac{1}{\tau_{\rm DM}} 
 \frac{\rho_{\rm halo}(\vec{r})}{m_{\rm DM}}
 \left[ \frac{dN_{e^+}}{dE} \right]_{\rm dec},
\end{eqnarray}
where $\tau_{\rm DM}$ is the lifetime of dark matter.  In the above
expressions, $[dN_{e^+}/dE]_{\rm ann}$ and $[dN_{e^+}/dE]_{\rm dec}$
are the energy distributions of the positron from single pair
annihilation and decay processes, respectively, and are calculated by
using PYTHIA package \cite{Sjostrand:2006za} for each dark matter
candidate.

\begin{figure}[t]
   \begin{center}
     \epsfxsize=0.45\textwidth\epsfbox{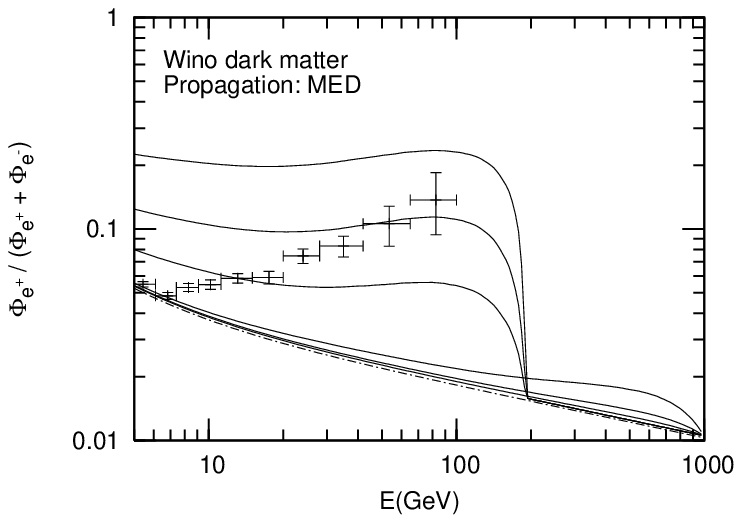}
     \epsfxsize=0.45\textwidth\epsfbox{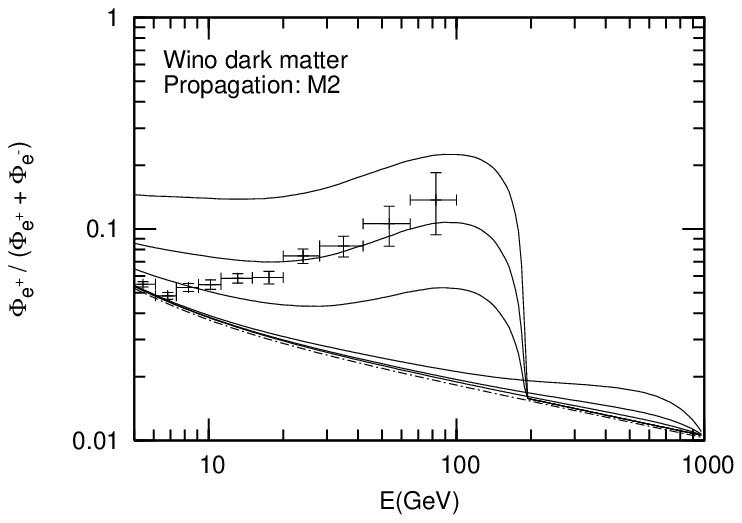}
     \caption{Positron fraction for the Wino dark matter case with
       MED (top figure) and M2 (bottom figure) propagation
       models.  We take $m_{\tilde{W}}=200\ {\rm GeV}$ (left) and $1\
       {\rm TeV}$ (right), and the boost factor is taken to be $B=1$,
       $3$, and $10$ (from bottom to top for each figure).  The
       dotted-dashed line is the positron fraction calculated only
       with the background fluxes.  The PAMELA data are also plotted.}
       \label{fig:wino}
   \end{center}
   \vspace{-0.5cm}
\end{figure}

Now, we discuss the positron flux for several models of superparticle
dark matter.  We first consider the case where the LSP is Wino-like
neutralino, which we denote $\tilde{W}^0$.  (Implications of the Wino
dark matter scenario to the PAMELA results were first discussed in
\cite{Cirelli:2008pk}.)  The thermal relic density of the Wino is much
smaller than the present mass density of dark matter if
$m_{\tilde{W}}\lesssim 1\ {\rm TeV}$ (with $m_{\tilde{W}}$ being the
Wino mass).  However, its mass density can be consistent with the
present dark matter density if the Wino is non-thermally produced in
the early universe \cite{Moroi:1999zb} (or if $m_{\tilde{W}}\simeq
2.5-3\ {\rm TeV}$ \cite{Hisano:2006nn}, which we do not consider in
this letter).  In addition, the Wino becomes the LSP in some of the
well-motivated scenarios of SUSY breaking, like the anomaly-mediation
model \cite{AMSB}.  If $\tilde{W}^0$ is dark matter, it mainly
annihilates into $W^+W^-$ pair when $m_{\tilde{W}}\lesssim 1\ {\rm
 TeV}$.  Then, the subsequent decays of $W^\pm$ produce energetic
positrons, which become the source of the cosmic-ray positron.

We calculate the positron flux from such an annihilation process.  In
Fig.\ \ref{fig:wino}, we plot the positron fraction using the MED and
M2 propagation models.  (Results with the M1 model are similar to
those with the MED model, and hence are not shown.)  In the figure,
the Wino mass is taken to be $200\ {\rm GeV}$, and $1\ {\rm TeV}$, for
which $\langle\sigma v\rangle=1.9\times 10^{-24}$, and $8.5\times
10^{-26}\ {\rm cm^3/sec}$, respectively.  In calculating the positron
fraction, we adopt the following background fluxes
\cite{Baltz:1998xv}: $[\Phi_{e^-}]_{\rm BG}=0.16 E_{\rm
GeV}^{-1.1}/(1+11E_{\rm GeV}^{0.9}+3.2E_{\rm GeV}^{2.15}) + 0.70E_{\rm
GeV}^{0.7}/(1+110E_{\rm GeV}^{1.5}+600E_{\rm GeV}^{2.9}+580E_{\rm
GeV}^{4.2})\ {\rm GeV}^{-1}\ {\rm cm}^{-2} \ {\rm sec}^{-1}\ {\rm
str}^{-1}$ for the electron, and $[\Phi_{e^+}]_{\rm BG}=4.5E_{\rm
GeV}^{0.7}/(1+650E_{\rm GeV}^{2.3}+1500E_{\rm GeV}^{4.2})\ {\rm
GeV}^{-1}\ {\rm cm}^{-2} \ {\rm sec}^{-1}\ {\rm str}^{-1}$ for the
positron.  As one can see, when the Wino-like LSP is dark matter, an
enhancement of the positron flux is possible.  However, irrespective
of the boost factor, such a scenario fails to realize the increasing
behavior of the positron fraction at $20\ {\rm GeV}\lesssim E\lesssim
100\ {\rm GeV}$ with the MED propagation model.  This is mainly
because the positrons from hadrons produced by the hadronic decays of
$W^\pm$ significantly contribute to the flux much below the threshold.
On the contrary, with the M2 model which tends to suppress low-energy
positrons relative to high energy ones, the shape of the positron
fraction becomes consistent with the PAMELA data at high energy region
(i.e., $E\gtrsim 20\ {\rm GeV}$) for $m_{\tilde{W}}\sim 200\ {\rm
GeV}$ and $B\sim 3$.  However, in such a case, the agreement between
the theoretical positron fraction and the PAMELA data is poor at
$E\lesssim 20\ {\rm GeV}$ with the present choice of background.  In
addition, in the Wino dark matter case, the anti-proton flux tends to
be too large \cite{Cirelli:2008pk}, which may exclude the possibility
of explaining the PAMELA positron excess in the Wino dark matter
scenario.

For a quantitative discussion, we calculate the $\chi^2$ variable
using the PAMELA data \cite{PAMELA}, neglecting effects of systematic
errors in the theoretical calculation.  Since the positron fraction in
the low-energy region is sensitive to the background fluxes, we only
use the data points with $E\geq 15\ {\rm GeV}$ ($5$ data points) in
the calculation of $\chi^2$.  (The detailed procedure to calculate
$\chi^2$ is the same as those given in \cite{Ishiwata:2008cu}.)  With
the M2 propagation model, the $\chi^2$ variable can become smaller
than $11.0$, which corresponds to the 95\ \% C.L. bound, in the
parameter region from $m_{\tilde{W}}=130\ {\rm GeV}$ (with $B\simeq
1$) to $m_{\tilde{W}}=310\ {\rm GeV}$ (with $B\simeq 10$).  In
addition, with MED and M1 models of propagation, the positron fraction
observed by the PAMELA is hardly realized with Wino dark matter
scenario with the conventional estimation of the background.

\begin{figure}[t]
   \begin{center}
     \epsfxsize=0.45\textwidth\epsfbox{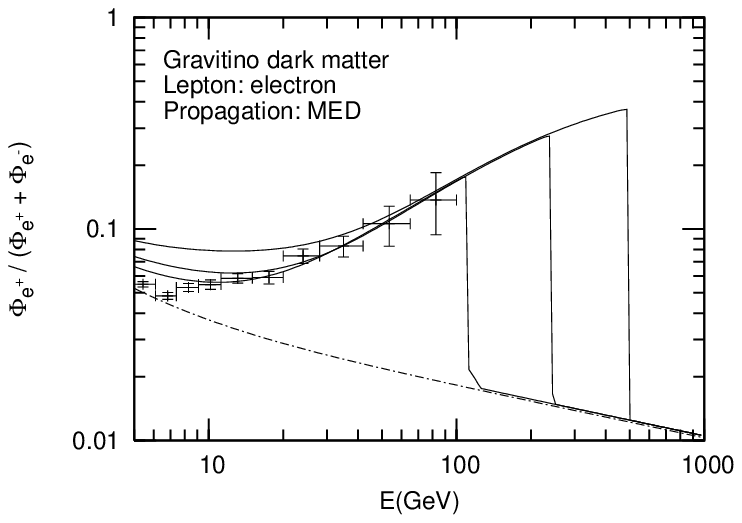}
     \epsfxsize=0.45\textwidth\epsfbox{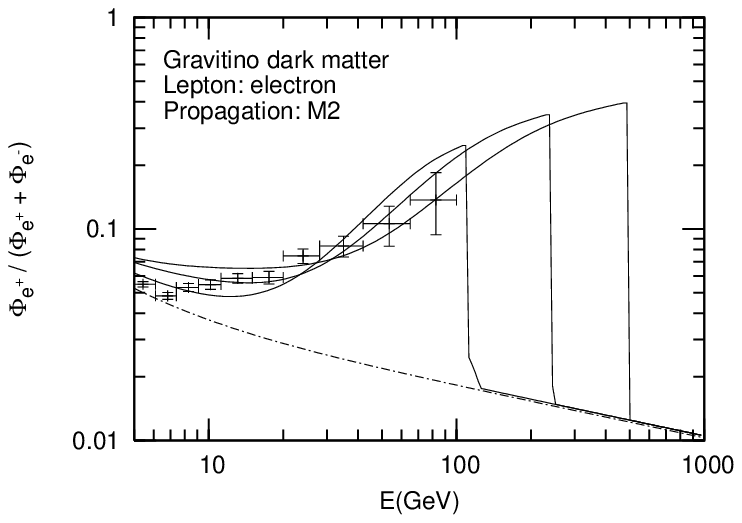}
     \caption{Positron fractions for the gravitino dark matter case.
       Gravitino is assumed to decay only into the first generation
       lepton (plus gauge or Higgs boson).  We take $m_{3/2}=250\ {\rm
         GeV}$, $500\ {\rm GeV}$, and $1\ {\rm TeV}$ (from left to
       right), and the MED (top figure) and M2 (bottom figure)
         propagation models are used.  For the case with the MED
         propagation model, we take $\tau_{3/2}= 8.5\times 10^{26}\
         {\rm sec}\times (m_{3/2}/100\ {\rm GeV})^{-1}$, while for the
         case with the M2 model, we take $\tau_{3/2}=1.8\times
         10^{26}\ {\rm sec}$, $9.0\times 10^{25}\ {\rm sec}$, and
         $6.3\times 10^{25}\ {\rm sec}$ for $m_{3/2}=250\ {\rm GeV}$,
         $500\ {\rm GeV}$, and $1\ {\rm TeV}$, respectively.}
     \label{fig:grav_el}
   \end{center}
   \vspace{-0.5cm}
\end{figure}

\begin{figure}[t]
   \begin{center}
     \epsfxsize=0.45\textwidth\epsfbox{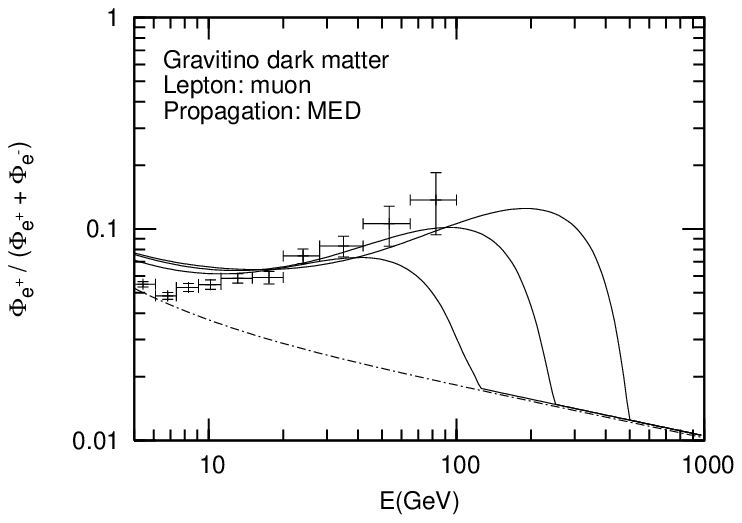}
     \epsfxsize=0.45\textwidth\epsfbox{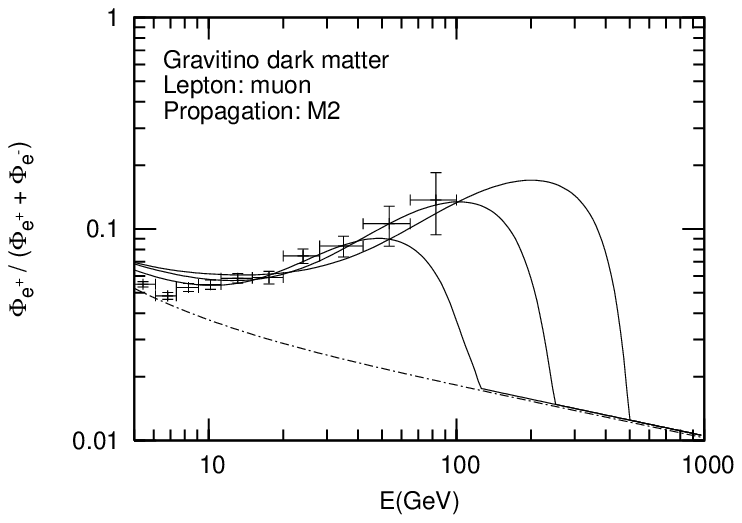}
     \caption{Positron fractions for the gravitino dark matter case.
       Gravitino is assumed to decay only into the second generation
       lepton.  We take $m_{3/2}=250\ {\rm GeV}$, $500\ {\rm GeV}$,
       and $1\ {\rm TeV}$ (from left to right), and the MED (top
       figure) and M2 (bottom figure) propagation models are used.
       For the case with the MED (M2) propagation model, we take
       $\tau_{3/2}=2.5\times 10^{26}\ {\rm sec}$, $1.6\times 10^{26}\
       {\rm sec}$, and $1.3\times 10^{26}\ {\rm sec}$ ($1.6\times
       10^{26}\ {\rm sec}$, $1.0\times 10^{26}\ {\rm sec}$, and
       $7.9\times 10^{25}\ {\rm sec}$) for $m_{3/2}=250\ {\rm GeV}$,
       $500\ {\rm GeV}$, and $1\ {\rm TeV}$, respectively.}
     \label{fig:grav_mu}
   \end{center}
   \vspace{-0.5cm}
\end{figure}

Next, we discuss the case where the gravitino (denoted as $\psi_\mu$)
is the LSP and hence is dark matter.  The pair annihilation cross
section is negligibly small in such a case.  However, with $R$-parity
violation (RPV), the gravitino LSP becomes unstable and energetic
positron can be produced by the decay.  Even if the gravitino is
unstable, it can be dark matter if the RPV is weak enough so that the
lifetime of the gravitino $\tau_{3/2}$ is much longer than the present
age of the universe \cite{Takayama:2000uz,Buchmuller:2007ui}.  In
fact, such a scenario has several advantages.  In the gravitino LSP
scenario with RPV, the thermal leptogenesis \cite{Fukugita:1986hr}
becomes possible without conflicting the big-bang nucleosynthesis
constraints.  In addition, the fluxes of the positron and $\gamma$-ray
can be as large as the observed values, and the anomalies in those
fluxes observed by the HEAT \cite{Barwick:1997ig} and the EGRET
\cite{Sreekumar:1997un} experiments, respectively, can be
simultaneously explained in such a scenario if $\tau_{3/2}\sim
O(10^{26}\ {\rm sec})$ \cite{Ishiwata:2008cu,Ibarra:2008qg}.

Here, let us consider the bi-linear RPV interactions.  Using the bases
where the mixing terms between the up-type Higgs and the lepton
doublets are eliminated from the superpotential, the relevant RPV
interactions are given by
\begin{eqnarray}
 {\cal L}_{\rm RPV} 
 = B_i \tilde{L}_i H_u + m^2_{\tilde{L}_i H_d} \tilde{L}_i H^*_d 
 + {\rm h.c.},
 \label{L_RPV}
\end{eqnarray}
where $\tilde{L}_i$ is left-handed slepton doublet in $i$-th
generation, while $H_u$ and $H_d$ are up- and down-type Higgs boson
doublets, respectively.  Then, the gravitino decays as
$\psi_\mu\rightarrow l_i^\pm W^\mp$, $\nu_i Z$, $\nu_i h$, and
$\nu_i\gamma$, where $l_i^\pm$ and $\nu_i$ are the charged lepton and
the neutrino in $i$-th generation, respectively.  Taking account of
all the relevant Feynman diagrams, we calculate the branching ratios
of these processes \cite{Ishiwata:2008cu}.  When the gravitino mass
$m_{3/2}$ is larger than $m_W$, the dominant decay mode is
$\psi_\mu\rightarrow l_i^\pm W^\mp$.  In addition, in such a case,
$\tau_{3/2}\simeq 6\times 10^{25}\ {\rm sec}\times
(\kappa_i/10^{-10})^{-2} (m_{3/2}/1\ {\rm TeV})^{-3}$, where
$\kappa_i=(B_i\sin\beta + m^2_{\tilde{L}_i
 H_d}\cos\beta)/m_{\tilde{\nu}_i}^2$ is the ratio of the vacuum
expectation value of the sneutrino field to that of the Higgs boson,
with $\tan\beta =\langle H^0_u \rangle / \langle H^0_d \rangle$, and
$m_{\tilde{\nu}_{i}}$ being the sneutrino mass.  Thus, the lifetime of
the gravitino is a free parameter and can be much longer than the
present age of the universe if the RPV parameters $B_i$ and
$m^2_{\tilde{L}_i H_d}$ are small enough.

We calculate the positron flux from the decay of the gravitino dark
matter.  For simplicity, assuming a hierarchy among the RPV coupling
constants, we consider the case where the gravitino decays selectively
into the lepton in one of three generations (plus $W^\pm$, $Z$, or
$h$).

When the gravitino decays only into first generation lepton, the
dominant decay mode is $\psi_\mu\rightarrow e^\pm W^\mp$ and a large
amount of positrons with the energy of $(m_{3/2}^2-m_W^2)/2m_{3/2}$
are produced by the decay.  Such monochromatic positron results in the
significant increase of the positron fraction at $20\ {\rm
  GeV}\lesssim E\lesssim 100\ {\rm GeV}$.  Then, contrary to the Wino
dark matter case, a drastic enhancement of the high energy positron
fraction is possible with any of the propagation model.  In such a
case, the PAMELA anomaly is well explained with the MED model; we
found that, with the M2 model, the high energy positron fraction is
too much enhanced to be consistent with the PAMELA data.  We found
that the $\chi^2$ variable may become smaller than the 95\ \%
C.L. bound (i.e., $11.0$) with the MED model; if $\tau_{3/2}$ is
properly chosen, the positron fraction well agrees with the PAMELA
data for $m_{3/2}\gtrsim 100\ {\rm GeV}$ irrespective of the gravitino
mass.  (Simultaneously, the energetic $\gamma$-ray flux is also
enhanced, which can be an explanation of the $\gamma$-ray excess
observed by the EGRET \cite{Sreekumar:1997un}.)  With the MED model,
we found that the the PAMELA data suggests $\tau_{3/2}\simeq 8.5\times
10^{26}\ {\rm sec}\times (m_{3/2}/100\ {\rm GeV})^{-1}$.  In Fig.\
\ref{fig:grav_el}, the positron fraction is shown with this choice of
the lifetime.  We can see a good agreement between the theoretical and
observational positron fractions.  Once the statistics will be
increased, the PAMELA may see the end-point at $E\simeq
(m_{3/2}^2-m_W^2)/2m_{3/2}$.  One can also see that the positron
fraction at $E\sim 10\ {\rm GeV}$ has a notable dependence on the
gravitino mass; with a better understanding of the background fluxes,
it may be used to derive a more stringent constraint on the gravitino
mass.

When the gravitino decays into second or third generation lepton, the
positron from the decay of the primary lepton (i.e., $\mu^+$ or
$\tau^+$) is less energetic than the positron directly produced by the
process $\psi_\mu\rightarrow e^+ W^-$.  Then, the energetic positron
flux becomes suppressed.  Even in such a case, it is still possible to
have an enhanced high energy positron flux.  For example, for the case
where the gravitino decays only into the second generation lepton, we
find the 95\ \% C.L. allowed region for any of the propagation model.
For the M2 model, which gives a better fit to the data than MED and
M1, the best-fit value of $\tau_{3/2}$ is $1.6\times 10^{26}\ {\rm
  sec}$, $1.0\times 10^{26}\ {\rm sec}$, and $7.9\times 10^{25}\ {\rm
  sec}$ for $m_{3/2}=250\ {\rm GeV}$, $500\ {\rm GeV}$, and $1\ {\rm
  TeV}$, respectively.  In Fig.\ \ref{fig:grav_mu}, we show the
positron fraction for these cases.  With the MED propagation model,
the agreement between the theoretical prediction and the PAMELA data
becomes slightly worse.

In the scenario with $\psi_\mu\rightarrow l^\pm W^\mp$, cosmic-ray
anti-proton is also produced.  However, we have checked that the
resultant anti-proton flux becomes comparable or smaller than the
observed values, taking account of the uncertainties in the
propagation model of anti-proton.  In particular, the best-fit
lifetime to explain the PAMELA anomaly is about 5 times longer than
that used in the calculation of \cite{Ibarra:2008qg}, and we obtain
the anti-proton flux smaller than that presented in
\cite{Ibarra:2008qg}.  The detailed analysis is given in
\cite{Ishiwata:2009vx}.

So far, we have concentrated on two important candidates for SUSY dark
matter.  Even in other cases, however, the positron flux may be also
enhanced.  For example, with the RPV interaction given in Eq.\
(\ref{L_RPV}), the PAMELA anomaly may be explained for the Bino dark
matter case.  In such a scenario, the Bino dominantly decays as
$\tilde{B}\rightarrow e^\pm W^\mp$ and the monochromatic positron
produced by the decay becomes the origin of the energetic cosmic-ray
positron.

\begin{figure}[t]
   \begin{center}
     \epsfxsize=0.45\textwidth\epsfbox{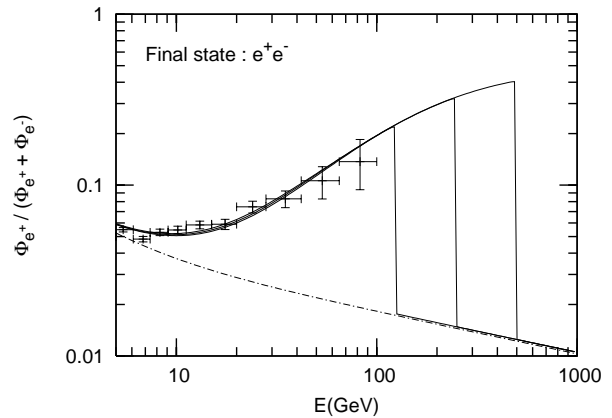}
     \caption{Positron fractions for the case where dark matter
       decays only into $e^+e^-$ pair.  We take $m_{\rm DM}=250\ {\rm
         GeV}$, $500\ {\rm GeV}$, and $1\ {\rm TeV}$ (from left to
       right), and $\tau_{\rm DM}=2.2\times 10^{27}\ {\rm sec}\times
       (m_{\rm DM}/100\ {\rm GeV})^{-1}$. The MED propagation model
       is used.}
       \label{fig:ee}
   \end{center}
   \vspace{-0.5cm}
\end{figure}

Our study indicates that the PAMELA anomaly may be well explained if
(almost) monochromatic positrons are emitted from the decay or the
pair annihilation of dark matter.  Thus, we finally consider the case
where the dark matter is unstable and dominantly decays into $e^+e^-$
pair.  This is the case if the left- or right-handed sneutrino is the
LSP and hence is dark matter, and also if the
$\hat{L}_i\hat{L}_1\hat{E}_1$ type RPV superpotential exists.  (The
left- and right-handed sneutrinos are viable candidates for dark
matter; see \cite{Hall:1997ah,Asaka:2005cn}.)  With the
$\chi^2$ analysis, we found that the positron fraction becomes
consistent with the PAMELA data when $\tau_{\rm DM}\simeq 2.2\times
10^{27}\ {\rm sec}\times (m_{\rm DM}/100\ {\rm GeV})^{-1}$ for $m_{\rm
 DM}\gtrsim 100\ {\rm GeV}$ in the MED model.  In Fig.\ \ref{fig:ee},
we show the positron fraction for $m_{\rm DM}=250\ {\rm GeV}$, $500\
{\rm GeV}$, and $1\ {\rm TeV}$, using the above $\tau_{\rm DM}$.  As
one can see, in such a case, an excellent agreement between the
theoretical prediction and the PAMELA data is obtained.

\noindent {\it Acknowledgements:} This work was supported in part by
Research Fellowships of the Japan Society for the Promotion of Science
for Young Scientists (K.I.), and by the Grant-in-Aid for Scientific
Research from the Ministry of Education, Science, Sports, and Culture
of Japan, No.\ 19540255 (T.M.).

\end{document}